# ImmunoLingo: Linguistics-based formalization of the antibody language

Mai Ha Vu[1,e,C], Philippe A. Robert[2,e], Rahmad Akbar[2,e], Bartlomiej Swiatczak[3], Geir Kjetil Sandve[4,J], Dag Trygve Truslew Haug[1,J], Victor Greiff[2,J,C]

[1] Department of Linguistics and Scandinavian Studies, University of Oslo, Norway
[2] Department of Immunology, University of Oslo and Oslo University Hospital, Norway
[3] Department of History of Science and Scientific Archeology, University of Science and Technology of China, China
[4] Department of Informatics, University of Oslo, Oslo, Norway
[e] Equal contribution
[J] Joint supervision
[C] Correspondence: m.h.vu@iln.uio.no, victor.greiff@medisin.uio.no

Highlights
- ➔ We propose ImmunoLingo, a linguistic formalization of antibody language to provide rigorous foundation for interpretable antibody language model design with the purpose of antibody specificity prediction
- ➔ Formalization builds on a precise understanding of the parallels between natural language and antibody sequences
- ➔ Formalized antibody language includes definition of well-formedness and meaning with characterizations of the lexicon and grammar
- ➔ To learn interpretable rules pertaining to antibody specificity only, the formalization suggests that input to antibody LMs contain structural information and semantics-based tokenization

## Abstract

Apparent parallels between natural language and biological sequence have led to a recent surge in the application of deep language models (LMs) to the analysis of antibody and other biological sequences. However, a lack of a rigorous linguistic formalization of biological sequence languages, which would define basic components, such as lexicon (i.e., the discrete units of the language) and grammar (i.e., the rules that link sequence well-formedness, structure, and meaning) has led to largely domain-unspecific applications of LMs, which do not take into account the underlying structure of the biological sequences studied. A linguistic formalization, on the other hand, establishes linguistically-informed and thus domain-adapted components for LM applications. It would facilitate a better understanding of how differences and similarities between natural language and biological sequences influence the quality of LMs, which is crucial for the design of interpretable models with extractable sequence-functions relationship rules, such as the ones underlying the antibody specificity prediction problem. Deciphering the rules of antibody specificity is crucial to accelerating rational and in silico biotherapeutic drug design. Here, we propose ImmunoLingo, a formalization of antibody language properties, and thereby establish not only a foundation for the application of linguistic tools in adaptive immune receptor analysis but also for the systematic immunolinguistic studies of immune receptor specificity in general.

# 1 Introduction

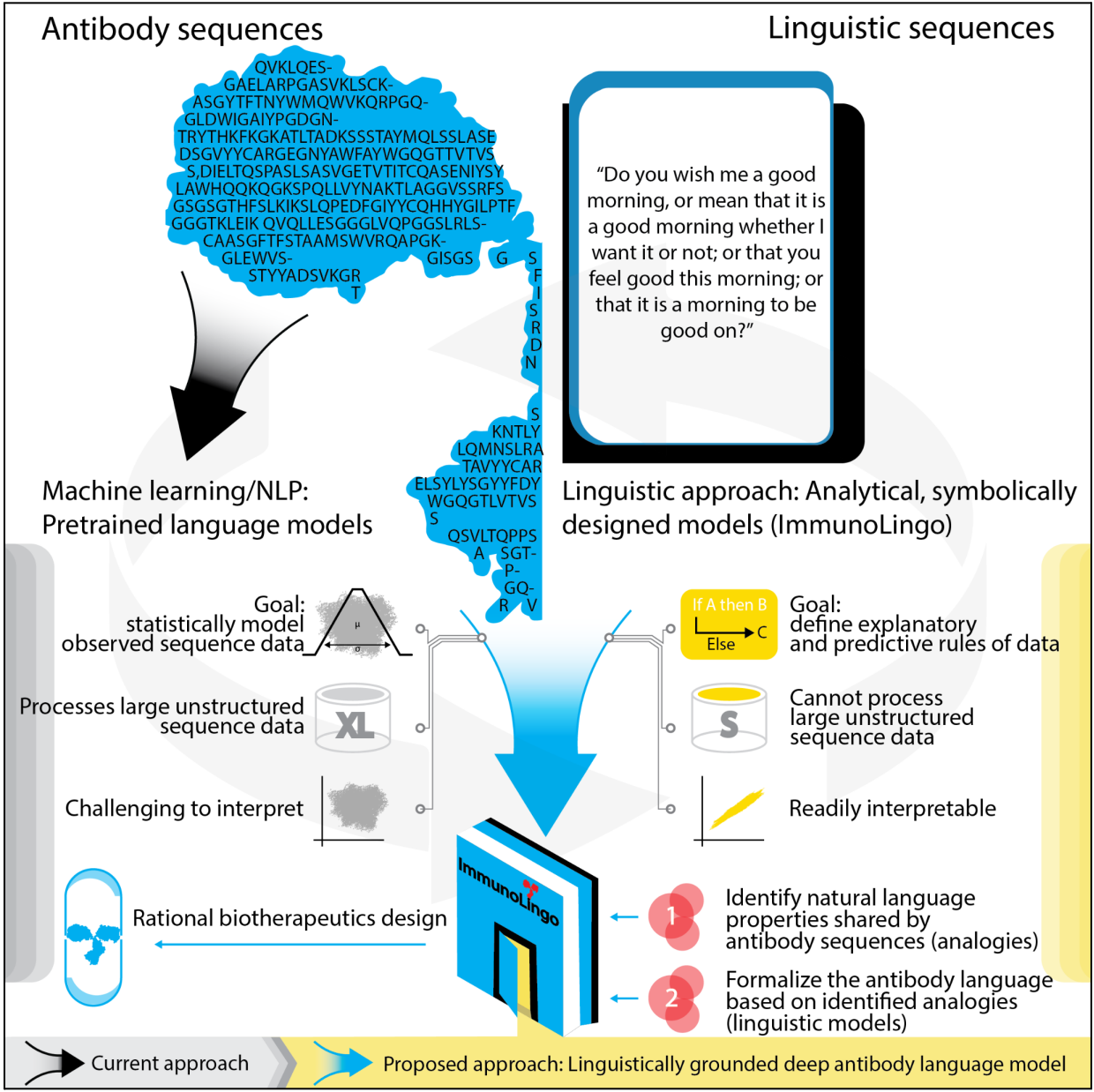

Figure 1 | **Integration of a linguistic formalization of antibody sequences (ImmunoLingo) into pre-trained language models leads to better interpretable antibody language models (LMs) with extractable sequence-function rules.** Apparent similarities between antibody and linguistic sequences (text from The Hobbit by J. R. R. Tolkien (Tolkien 1937)) motivate the application of language analysis tools to biological sequences, such as pretrained, statistical language models (LMs) (ML/NLP approach, left) on the one hand and analytical, rule-based, symbolic models on the other hand (linguistics approach, right). Pretrained antibody LMs are currently surging in popularity, but lack grounding in a rigorous linguistic formalization hindering the development of readily interpretable models with extractable rules. Here we propose a two-step linguistic formalization of the antibody language (ImmunoLingo) to guide interpretable antibody LM design: (1) Identify natural language properties shared by antibody sequences (analogies), and (2) formalize the antibody language based on the identified analogies (linguistic models). Integrating a linguistic formalization of antibody sequence language into pretrained antibody LMs can lead to better interpretability while maintaining the ability to statistically process large, unstructured data,

resulting in a linguistically grounded deep antibody LM. Easily interpretable LMs can aid with deciphering the rules of antibody specificity, which is crucial for rational and in silico antibody biotherapeutics design.

DNA and protein code, like human language, constitute a system of sequences: information-conveying strings of units (Russell 1931; Beadle and Beadle 1966; Gimona 2006; Searls 2002). Antibody sequences in particular display complex interactions that are suggestive of an underlying nontrivial grammar. This parallel between natural language and biological sequences has warranted applications of shared analytical methods for their study, most recently Natural Language Processing (NLP) tools like deep, self-supervised language models (LMs) (Bepler and Berger 2021; Ofer, Brandes, and Linial 2021; Alley et al. 2019; Brandes et al. 2022; Elnaggar et al. 2021; Heinzinger et al. 2019; Rao et al. 2019; Hie, Yang, and Kim 2022; Unsal et al. 2022; Rives et al. 2021; Meier et al. 2021; Wang et al. 2019; Xu et al. 2022; Nijkamp et al. 2022; Flam-Shepherd, Zhu, and Aspuru-Guzik 2022; Vu et al. 2022).

However, popular NLP tools have been applied without grounding in a rigorous linguistic formalization of biological sequences, which we argue is necessary to uncover biological sequence grammar, i.e. interpretable finite rules over discrete units that govern biological sequence well-formedness and function (Vu et al. 2022). That such rules exist in biological systems cannot be doubted, as combinations of interactions, domains, motifs and folds depend on defined thermodynamic and kinetic principles of molecular assembly and molecular interaction (Searls 2002; 2010; Gimona 2006; Jumper et al. 2021). Explicit knowledge of biological sequence rules can further aid rational therapeutics design (e.g., antibody therapeutics), as it can guide and limit the search for target-specific and developable sequences (Akbar et al. 2022).

A linguistic formalization, which would involve defining the lexicon and the grammar (both well-formedness and meaning-related rules) of a given biological sequence language, can provide a more precise understanding of the parallels between linguistic and biological sequences, and thus enable a more domain-adapted application of LMs. Defining the meaning of a biological sequence is relevant for protein-protein interactions because a possible definition of protein sequence meaning is the bound structures. Antibody sequence specificity then is of particular interest for a linguistic approach, as it is an even more complex problem than that of non-immune protein-protein binding interaction (Akbar et al. 2022; 2021). Furthermore, a challenge to ML models is the accurate prediction of out-of-distribution data (Teney, Oh, and Abbasnejad 2022), whereas the stated purpose of linguistic grammars is the ability to model all possible sequences regardless of statistical frequency (Chomsky 1964).

Here we propose two main points to consider when linguistically formalizing biological sequences (Figure 1): (i) identify linguistic features in biological sequences to enable deep understanding of the possibilities and limits of the linguistic analogy (**analogies**) and (ii) formalize the language of the biological sequence with rigorously defined characterization of its components to enable the design of a theoretically grounded deep language model (**linguistic models**). Our proposal for formalizing biological sequences as language aims to be flexible guidance for a rigorous and thoughtful linguistic formalization that can serve as grounding for NLP tool applications to biological sequences. The concrete implementation of the steps outlined here will depend on the particular type of biological sequence and the research question. In this paper, we propose ImmunoLingo, a possible implementation of our proposal on formalizing the antibody language in the context of the antibody specificity prediction problem**.** Our suggestions are meant as guidance for future directions addressing challenges in antibody specificity prediction with current NLP tools.

## 2 Formulation of the antibody specificity prediction problem

Antibodies are proteins produced by the immune system to bind and support the clearance of foreign molecules in the body. It has been known for a century now that antibodies can recognize virtually any given natural and non-natural structure, proteic, lipidic, or even glycan. Structures recognized by antibodies are called antigens (Landsteiner 1945). Antibody function is mediated by binding to an antigen. The receptor (binding interface) determines the strength of interaction and biological function. In the case of protein antigens, the paratope and epitope refer to antibody, and respectively antigen, residues involved in the binding interface (usually defined as residues within a distance of 5Å of one another (Akbar et al. 2021)).

One of the longest-standing problems in immunology is the antibody specificity prediction problem, which can be formulated as follows: given an antibody, enumerate all potentially binding proteins (antigens) and vice versa: given an antigen, enumerate all potentially binding antibodies (Robert et al. 2021). One of the greatest challenges regarding antibody specificity prediction is cross-reactivity (Robert, Marschall, and Meyer-Hermann 2018; Boughter et al. 2020; Neumeier et al. 2022; Cunningham et al. 2021). Antibody-antigen binding is a many-to-many mapping where the same immune receptor can recognize multiple different antigens, and an antigen can be recognized by multiple antibody sequences (Mason et al. 2021; Robert et al. 2021; Robert, Marschall, and Meyer-Hermann 2018; Greiff, Yaari, and Cowell 2020), creating a complex recognition network between antibodies and antigens. Prediction of antibody specificity is further complicated by the fact that sequences with close edit distance may bind different sets of antigens, while dissimilar sequences can bind the same antigens (Robert et al. 2021; Akbar et al. 2022; Mason et al. 2021; Sangesland et al. 2022).

While the connections between language and adaptive immunity have been previously discussed to some extent (Focus Box 1), this work is the first to our knowledge to map linguistic concepts rigorously and systematically to antibody sequence properties, with specific attention to antibody specificity prediction.

Focus Box 1 | Early endeavors in immunolinguistics

One of the earliest works connecting linguistics and immunology can be traced back to (Burnet 1972), who drew linguistic analogies to theorize about the origin of immune receptor variation. He compared immune receptors to short strings of letters that are randomized by a computer program (Burnet 1972, p. 39-40). He speculated that mechanisms might be in place to increase the frequency of meaningful combinations of stochastically generated letter-like gene fragments during development, which has been experimentally supported later (Krovi et al. 2019).

A more elaborate version of the linguistic framework was provided by Niels Jerne in his 1984 Nobel lecture entitled "Generative grammar of the immune system" (Jerne 1985). Jerne drew a parallel between the open-endedness of language, which refers to language's creative capacity to express any possible semantic meaning (Chomsky 1964; 2009) and the completeness hypothesis of the antibody repertoire, which proposes that every possible foreign antigen can be recognized by a pre-existing antibody sequence in the organism (Perelson 1989; Coutinho 1980). By comparing immune receptor specificity to a linguistic meaning and highlighting the importance of innate learning skills in language acquisition, he alluded to universal combinatorial rules of immunity akin to those postulated by Chomsky in language.

These early insights were further developed by linguists who saw in the immune system an example of a system that, despite being creative and productive is also innate and conserved, reliant on inborn rules and structures (Piattelli-Palmarini 1986; Piattelli-Palmarini and Uriagereka 2004). Inspired by advances in the studies of antibody specificity, these authors suggested that selective rather than instructive (Lamarckian) mechanisms mediate language acquisition process.

The semantic dimension of immunity was analyzed, in turn, jointly by semioticians (specialists in the study of meaning and signs) and immunologists, who attempted to frame intercellular immune exchanges in terms of symbolic communication and meaning-making (Sercarz et al. 1988). As an extension of these early efforts, Atlan and Cohen considered the process of information processing in the immune system, suggesting that immune meaning unfolds in a complex process following receptor activation (Atlan and Cohen 1998). Although bearing promise to expand the conceptual framework of immunology and provide novel testable hypotheses in the field, these early immunolinguistic considerations were largely speculative, as supporting large-scale Ig sequencing data were unavailable at that time.

## 3 Analogies: Identifying properties of natural language in antibody sequences

The prerequisite to treating antibody sequences as a language is the existence of analogies, and these analogies provide the foundation for ImmunoLingo. Here, we identify the following key properties of natural language present in antibody sequences: (1) discreteness (Hockett 1960), (2) hierarchical structure (Chomsky 1956), and (3) ambiguity (Hindle and Rooth 1993; Guo 1997) (Figure 2)**.** Additionally, (4) semantic compositionality (Montague 1970), which is essential for explaining linguistic ambiguity, might also apply to antibody sequences.

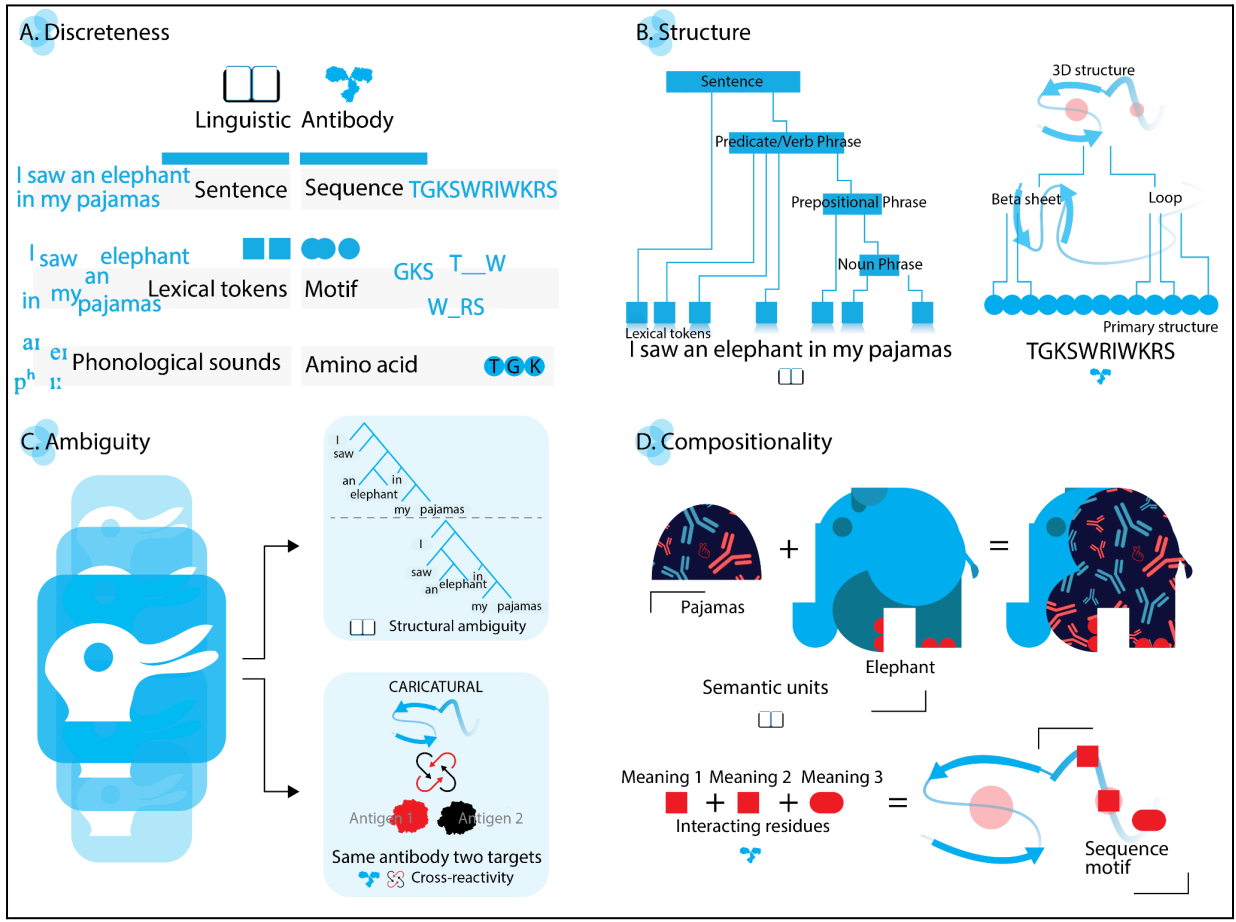

Figure 2 | **Antibody sequences display the linguistic properties (A) discreteness, (B) hierarchical structure, (C) ambiguity, and (D) compositionality.** (A) Sentences are built from finite building blocks (sounds) via intermediary units (words/lexical items/tokens) that possess semantic meaning. Antibodies similarly are built from a finite set of amino acids and likely also possess intermediary units (motifs). (B) Linguistic strings possess a combinatorial building structure that can be represented as trees following a set of syntactic rules. Similarly, antibodies have a three-dimensional structure, which results from the folding of sub-sequences into primary structure blocks, that can only form a finite set of 3D structures according to physical constraints. (C) Ambiguity, the idea that the same structure can map to two distinct meanings (e.g., the picture on the left can be either a rabbit or a duck), is found in both linguistic and antibody sequences: natural language sentences can have multiple meanings, and antibody sequences can bind multiple targets (cross-reactivity). (D) The meaning of a sentence is composed by combining the meaning of its lexical items that build it through the order they are combined in the sentence. We hypothesize that antibody meaning can be similarly composed from individual motifs with associated meaning, thus accounting for both structural and tokenization ambiguity.

## 3.1 Shared property 1: Discreteness

Linguistic sequences are built from a finite number of smaller units into a countably infinite number of possible combinations with the help of a finite set of rules (Hauser, Chomsky, and Fitch 2002): sounds are combined into lexical items, lexical items are combined into phrases, and phrases are built into sentences. At every step, language-specific rules determine an infinite number of possible combinations due to the possibility for recursive rules (Hauser, Chomsky, and Fitch 2002).

Antibody sequence data can also be subdivided into smaller units, but the challenge is to find the level of discrete units that can give rise to useful interpretable rules for predicting antibody specificity. The one obvious level of abstraction is on the level of amino acids, which are most analogous to sounds in language (Figure 2A): both amino acids and sounds are well-identifiable, small units that by themselves do not bear a more abstract, functional meaning. Antibody sequences, like all proteins, are composed of amino acids, and there is a possibility of $10^{14}$ combinations of these amino acids for recombined antibodies, but only $10^5$–$10^6$ for non-immune proteins (Greiff et al. 2017; Elhanati et al. 2015; Tress, Abascal, and Valencia 2017). At a higher level, there is also reason to believe that there exist meaningful units that are larger than amino acids, analogous to lexical items in language. Subdividing an antibody sequence into intermediate units of a size that ranges between that of individual amino acids and the full sequence was suggested to lead to improved prediction accuracy of antigen binding (Akbar et al. 2021; Robert et al. 2021; Pantazes et al. 2016). Other studies have also designed de novo stable antibody sequences by combining protein fragments (Aguilar Rangel et al. 2022; Zhou, Panaitiu, and Grigoryan 2020).

## 3.2 Shared property 2: Hierarchical structure

That language is organized in hierarchical structures has been proposed by linguists for decades (Chomsky 1956). Linguistic sentence structure is built with a series of rules and is commonly modeled as a tree (Figure 2B).

Antibodies have 3D structure beyond their linear sequence (Figure 2B), similarly to proteins. Protein sequences fold into a primary and secondary structure (Figure 2B) with local structural patterns defined by the local amino acid sequence (which could be seen as atomic lexical units), followed by the 3D conformation of longer-distance domains (tertiary and quaternary structure, which could be seen as structure involving larger units such as phrases) (Rossmann and Argos 1981; Qing et al. 2022). The order of local folding events with appearance of local secondary structures before the global folding of the protein can, for instance, be represented as a tree of folding events (Hockenmaier, Joshi, and Dill 2006; Searls 2013).

The 3D structure of an antibody directly contributes to its specificity: the folding creates paratopes on the antibody (Sela-Culang, Kunik, and Ofran 2013). Of note, the same antibody sequence may bind different targets using a different set of paratope residues (Boughter et al. 2020; Bunker et al. 2017; Lecerf et al. 2019), and consequently, non-paratope parts of antibodies also contribute to the specific binding by making the appropriate 3D structure possible (Sela-Culang, Kunik, and Ofran 2013).

## 3.3 Shared property 3: Ambiguity

Linguistic sequences can be ambiguous, as they can map to multiple different meanings (Figure 2C). For example, the sentence "I saw an elephant in my pajamas" displays structural ambiguity (Hindle and Rooth 1993): depending on the underlying structure, it either means that the speaker wears their own pajamas or that the elephant wears the speaker's pajamas (Figure 2C). Other possible types of ambiguities include tokenization ambiguity (i.e., the string sequence can be subdivided into different sets of tokens (Guo 1997)) and lexical ambiguity (i.e., the same token can have multiple meanings).

Antibodies display ambiguity through cross-reactivity, which can further be categorized as polyreactivity (recognition of unrelated antigens with different epitopes), promiscuity (recognition of several mutated variants), and conserved recognition (recognition of unrelated antigens with the same conserved epitope region) (Robert, Marschall, and Meyer-Hermann 2018). Both polyreactivity and promiscuity fit the notion of linguistic ambiguity. On the other hand, conserved recognition does not, because the antibody sequence itself does not bind different epitope structures; this makes conserved recognition more similar to mapping the same phrase to the same meaning in various different larger contexts. Discussing cross-reactivity in terms of linguistic ambiguity thus stays relevant if polyreactivity and promiscuity account for a relatively large portion of antibody cross-reactivity cases compared to conserved recognition, but exact percentages of the various categories of cross-reactivity remain unknown.

As with language, antibody sequence ambiguity could be due to structural ambiguity, where each different antibody fold associated with a given sequence binds different targets (Fernández-Quintero et al. 2019) (Figure 3B). However, often the same antibody structure can also recognize several different antigens (Robert, Marschall, and Meyer-Hermann 2018; Cunningham et al. 2021; Boughter et al. 2020), where tokenization or lexical ambiguity might become relevant: either the same antibody sequence could be re-analyzed as containing different tokens or the same token has multiple associated functions (Figure 3B).

### 3.4 Shared property 4: Compositionality of meaning

Every linguistic sequence maps to meaning, which refers to concepts that are grounded outside of language. The meaning of a lexical item is arbitrary, as it is not encoded in the letters or sounds that built it; for example, nothing in the spelling or pronunciation of the word "elephant" indicates its meaning. On the other hand, the meaning of sentences is compositional, as it can be derived from the meanings associated with the individual lexical items that built it and from the order they combined.

Compositionality is a key feature of natural language for explaining ambiguity. The two meanings for "I saw an elephant in my pajamas'' (Figure 2D) is due to the different orders in which the lexical items in the sentence combined to arrive at different meanings. The phrase ''in my pajamas" can combine directly either with the noun "elephant" or the verb "saw". Combining with the meaning of "an elephant" results in the meaning of "an elephant in my pajamas", deriving the first meaning (Figure 2D, left), whereas combining with "saw an elephant", the result is the second meaning of the sentence (Figure 2D, right).

Antibody binding is governed by complex physicochemical laws, which in principle makes antibody meaning nonarbitrary. However, in practice where only sequence information is given, it is unclear to what extent antibody sequence meaning is arbitrary (behaving more like lexical items) rather than compositional (behaving more like sentences). Based on evidence so far, it is likely that even in this case, antibody meaning is not completely arbitrary, and there is predictive correspondence between the amino acids of antibody sequences and the antigens they bind. Indeed, multiple recent works have shown that antigen-binding (and even affinity) can be predicted based on the amino acid sequence alone (Mason et al. 2021; Bachas et al. 2022; Makowski et al. 2022). These works are complemented by first successes in paratope-epitope prediction (Pittala and Bailey-Kellogg 2020; Jespersen et al. 2019; Akbar et al. 2021; Del Vecchio et al. 2021). More generally, since antigen specificity is mainly determined by ≈5–20 amino acids, the recognition and classification of billions of different antigens is based on decision boundaries within a feature space of merely ≈5–20 dimensions. The low dimensionality implies the presence of strong high-order dependencies between amino acids (dimensions). Consequently, as previously explained in (Brown et al. 2019), antigen specificity may result from the conditional combination of components in the form of paratope subsequences (contiguous or gapped k-mers), similar to natural language where semantics arise from a combination of words according to a given grammar. Indeed, Akbar et al.'s study showed that a systematic subdivision of antibodies into interacting and non-interacting paratope motifs improved antigen binding prediction (Akbar et al. 2021), and importantly, these motifs were shared across entirely different antibody-antigen binding complexes suggesting the existence of a generalizable, systematic antibody grammar. If antibody meaning can be derived compositionally, then identifying the compositional rules governing binding specificity is the key to solving the antibody specificity prediction problem.

## 4 Linguistic model: Formalizing the antibody language based on identified linguistic analogies

Here we describe ImmunoLingo, our formalization of the antibody language and characterize its components. The formalization builds on the shared properties between natural language and antibody sequences (discreteness, hierarchical structure, ambiguity, and compositionality of meaning), while also considering differences between the two systems.

### 4.1 Defining well-formedness and meaning for the antibody language

To formalize the antibody language with respect to answering the antibody specificity problem, it is first necessary to define what constitutes a well-formed structure and its meaning (Saussure 1986). For a given natural language, a sequence or structure is well-formed if it adheres to the language's well-formedness rules (phonological rules for well-formed sound sequences, i.e. lexical items, and syntactic rules for well-formed lexical item sequences, i.e. sentences), and each syntactically well-formed sentence maps to meaning with the help of compositional semantic rules.

We propose that well-formed antibody sequences are the set of all observable antibody sequences, i.e, those sequences formed by V(D)J recombination that are in-frame and without stop codons. (Figure 3A). Syntactic rules for natural language build well-formed sentence structures directly (Chomsky 1956), and well-formed sentences can be read off the structure. In contrast, for antibody sequences, we distinguish between sequence-building syntactic rules, which determine well-formed one-dimensional sequences on the one hand and structure-building syntactic rules, which map well-formed sequences to well-formed three-dimensional folded structures on the other hand. An example of sequence-building rules is V(D)J recombination, the pseudo-random process of recombining germline V, D, and J segments to initially generate a diverse set of adaptive immune cells (Hozumi and Tonegawa 1976), and an example of structure-building syntactic rules would be the rules that determine antibody folding.

Following the analogy between linguistic ambiguity and antibody polyreactivity and promiscuity, the meaning of a well-formed antibody sequence could be defined as the set of epitopes it binds (Figure 3A). In natural language, compositional semantic rules map structure to meaning. Analogously, antibody semantic rules map from folded antibody structure to its recognized epitopes. If the input to antibody specificity prediction is an already well-formed (i.e., observed) antibody sequence, but not the fully folded structure, deciphering both the structural syntactic rules and the compositional semantic rules becomes key to solving the antibody specificity prediction problem. However, these two types of rules can be teased apart if the input is a well-formed structure. The rules enable the rational mapping from antibody sequence to recognized epitopes, which may be further linked to full antigen structures.

## 4.2 Antibody language has separate syntactic and semantic lexicons containing discrete units for syntactic and semantic rules

Both syntactic and semantic rules operate over sets of discrete units; these are the syntactic and semantic lexicons, respectively. While it is a requirement for the semantic lexicon that its items have functional meaning that can be added together into a combined meaning, the same is not necessarily true for the syntactic lexicon, since syntactic rules only determine if a certain combination of items leads to a well-formed sequence or structure. While in natural language syntactic and compositional semantic rules share the same lexicon, the same is not true for antibody sequences. Subsequently, our formalization defines separate syntactic and semantic lexicons for the antibody language.

The antibody semantic lexicon consists of lexical items (i.e., discrete units that correspond to biological motifs with an identifiable functional meaning) that cannot be subdivided further into smaller components with meaning (Figure 3A illustrates a few hypothetical examples, where motifs possibly correspond to psychochemical properties, shape, and content of corresponding antigen epitope). Because compositional semantic rules map from structure to meaning, the semantic lexical items should already contain knowledge of motif structure (Figure 3A). As with linguistic lexical items, these motifs can be lexically ambiguous with multiple different meanings, and multiple motifs could be synonymous by mapping to the same meaning as well. Furthermore, antibody motifs might overlap, be non-contiguous, and a sequence might not break down neatly into separate motifs; there could be remaining single amino acids that do not belong to any particular motif and are only required for well-formedness reasons. To develop a full semantic lexicon of antibodies, there needs to be an exhaustive list of relevant lexical meanings to antibody specificity prediction and an analytical mapping between the lexical meaning to the corresponding motifs. Compositional semantic rules iteratively map the combinations of functional motifs to a combined meaning, resulting in the features of a full recognized epitope (Figure 3A, semantic rules).

In contrast, the antibody syntactic lexicon does not necessarily consist of meaningful lexical items. For example, V(D)J recombination operates on the level of nucleotides, which do not have identifiable functional meaning that pertain to antibody specificity, and thus are unlikely to be the appropriate level of discrete units for compositional semantic rules. (Figure 3A, syntactic rules). Other possible syntactic rules might operate over individual amino acids or even motifs, but in neither case do they necessarily have a corresponding functional meaning; they are useful as syntactic units if they can facilitate the emergence of interpretable syntactic rules (Figure 3A, syntactic rules). The syntactic lexicon is only relevant for antibody specificity prediction in so far as it is relevant for structural syntactic rules, and facilitates finding compositional semantic rules.

## 4.3 Linguistic perspective on challenges for antibody specificity prediction

Here we show in detail how the challenges for antibody specificity prediction are addressed in our linguistic formalization. In particular, we discuss the lack of correlation between sequence similarity and binding specificity (Robert et al. 2021; Akbar et al. 2022; Mason et al. 2021; Robinson et al. 2021), as well as cross-reactivity (Robert, Marschall, and Meyer-Hermann 2018).

Sequence similarity is dependent on the tokens that are chosen for calculating edit distance. For example, "cat" and "car" are similar sequences in a letter-based tokenization, but not in a word-based tokenization. In biology, sequence similarity is typically measured based on amino acid edit distance, and thus antibody sequences that are only a few edit distances away in terms of amino acids are considered similar. However, amino-acid based similarity is a poor indicator for binding, as similar sequences can bind different antigens, and dissimilar sequences can bind the same antigen (Robert et al. 2021; Akbar et al. 2022; Mason et al. 2021; Robinson et al. 2021). The linguistic formalization suggests that the relevant tokens to antibody specificity instead should be based on a semantic lexicon. With semantic tokens forming the basis for calculating sequence similarity, there could be a stronger correlation between sequence similarity and specificity.

Antibody cross-reactivity can be understood as linguistic ambiguity (Figure 3B). Antibody sequences display structural ambiguity by having structural syntactic rules that allow multiple different folds, and therefore different paratopes. Each fold maps to different sets of epitopes, similarly to how different sentence structures mapped to different meanings in natural language (Figure 2C).

Antibody sequences also display tokenization ambiguity when the compositional semantic rules analyze the same antibody structure as built from different sets of antibody motifs (Figure 3B). For example, given the structure CAR_IC_AT_URAL_ in Figure 3B (underlined amino acids signal interacting motifs) and a semantic lexicon (Figure 3A), the antibody could be analyzed with different semantic tokenizations: either as containing the semantic lexical items I_C_*U_ + RA_*_ or I_ + C_AT + U_*L_. Each tokenization maps to a different combined meaning, and thus different sets of epitopes.

Finally, lexical ambiguity is also possible, if each lexical item can be associated with multiple meanings. Then even the same tokenization might further map to different possible epitopes (Figure 3B).

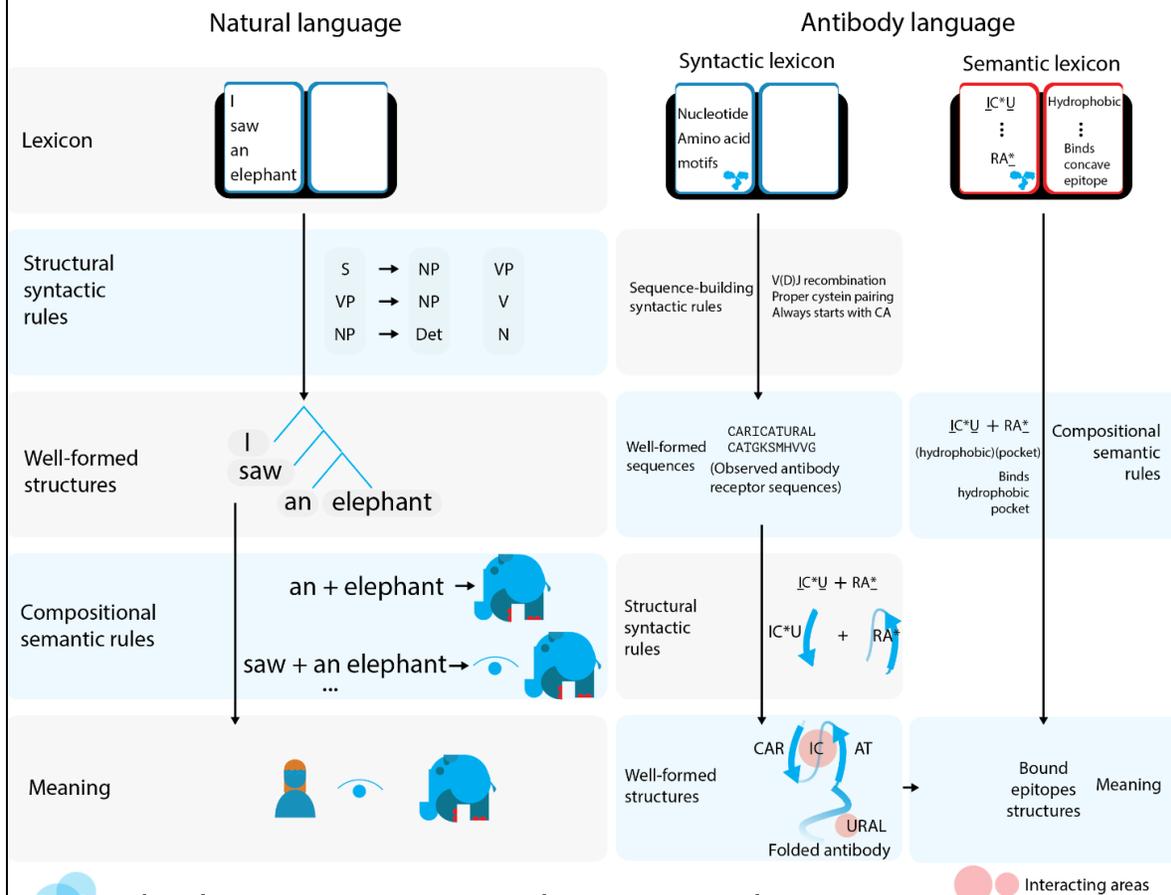

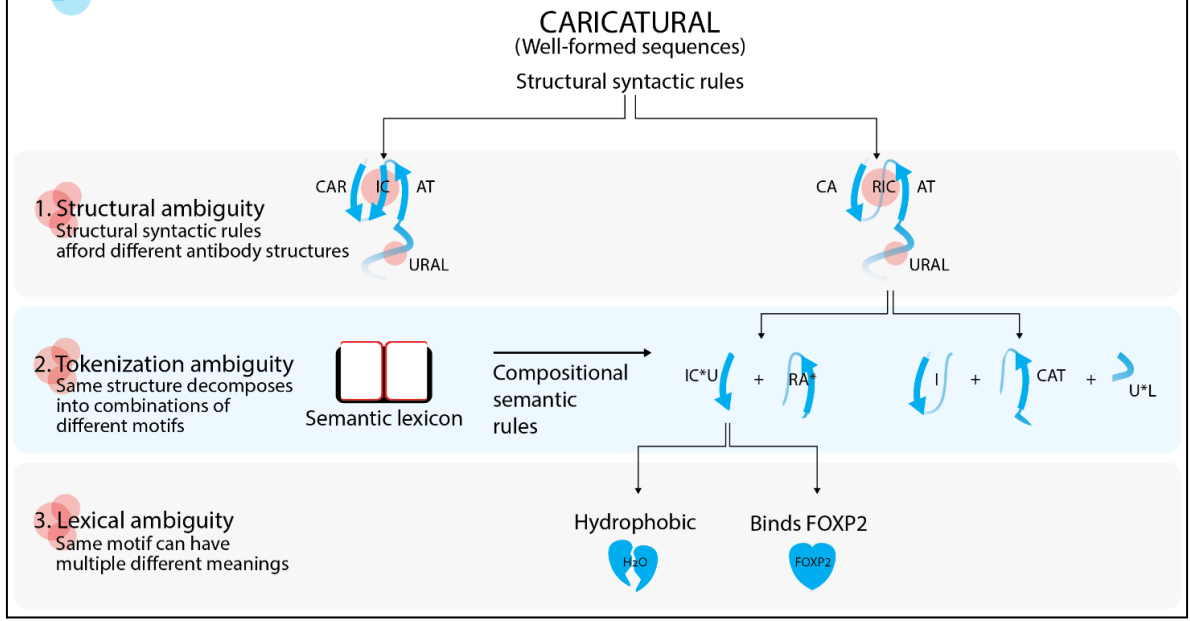

Figure 3. | (A) **ImmunoLingo: The formalization of the antibody language.** Natural language builds well-formed structures from the lexicon with the use of structural syntactic rules, and compositional semantic rules map structures to meaning. In the antibody language, on the other hand, sequence building syntactic rules (e.g., V(D)J recombination) build well-formed sequences from the syntactic lexicon (e.g., nucleotides, amino acids, motifs), and structural syntactic rules map well-formed sequences to well-formed structures. Semantic compositional rules draw from a separate semantic lexicon to map folded antibody structures to meaning, i.e. bound epitopes. (B) **Antibody cross-reactivity as linguistic ambiguity.** Antibody cross-reactivity can be analyzed as different types of linguistic ambiguity. (1) Structural ambiguity arises when structural syntactic rules allow different folded antibody structures, and each structure binds different sets of epitopes. (2) Tokenization ambiguity arises when the same structure can be subdivided into combinations of different motifs in the semantic lexicon. (3) Lexical ambiguity arises when the same motif can have multiple different meanings.

## 5 Discussion: Linguistic formalizations guide antibody LM design with the goal of gaining interpretability

Here we discuss how the formalization of antibody language provided by ImmunoLingo can guide antibody language model (LM) design. Pre-trained LMs are self-supervised on unlabeled sequence data (pre-training data) and hence are probabilistic models over sequences of tokens (Qiu et al. 2020). Current antibody LMs (Leem et al. 2021; Olsen, Moal, and Deane 2022; Ruffolo, Gray, and Sulam 2021; Shuai, Ruffolo, and Gray 2021; Ruffolo, Sulam, and Gray 2022; Prihoda et al. 2022; Ostrovsky-Berman et al. 2021) typically use amino acid or at most k-gram based tokenization with focus on prediction accuracy rather than interpretability and explainability. However, to gain useful information for rational antibody design applications, the rules extracted from antibody LMs with the help of various interpretability methods (Rogers, Kovaleva, and Rumshisky 2021; Vig et al. 2021; Weiss, Goldberg, and Yahav 2020) should be easily understandable for human scientists. The linguistic formalization clarifies that deciphering the compositional semantic rules are the key to directly answer the antibody specificity prediction problem, and thus a future goal could be to build antibody LMs dedicated to learning the semantic rules for easy interpretability and explainability. Since all information not explicitly provided by the input will have to be latently learned during self-supervised training, there needs to be careful consideration of how pre-training data selection, data encoding, and tokenization influence the rules learned by the LM.

The linguistic formalization further reveals that there is a possible disconnect between the syntactic and semantic lexicon, which can lead to worse antibody LM performance for antibody specificity recognition, compared to natural language LM performance on semantic tasks. The syntactic lexicon can be inferred from distributional properties of the sequence, therefore an LM can learn to recognize syntactically well-formed sequences and structures through self-supervised training alone. Because for natural language, the syntactic lexicon is the same as the semantic lexicon, knowing the syntactic lexicon is adequate for performing semantic tasks. For antibody sequences, on the other hand, the disconnect between syntax and semantics means that to perform "semantic" tasks (i.e., antibody specificity prediction), the antibody LM needs to separately learn the semantic lexicon, which cannot be simply inferred from sequence alone.

Since the aim of the LM is to learn an accurate representation of the language to be modeled, the pre-training data should reflect that language. According to linguistic formalization, the language consists of the set of all well-formed sequences (or structures). Thus, for learning antibody language, the pre-training data needs to be representative of all well-formed (i.e., observed) antibody sequences so that the model does not learn a skewed pattern.

If the input is unannotated data of all observed antibody sequences, the LM will have to learn both structural syntactic rules and semantic mapping rules in order to perform antibody specificity prediction. Unless it is known which parts of the model capture different types of rules (similarly to what is known for natural language BERT architecture (Tenney, Das, and Pavlick 2019)), the extracted rules from the model might be too complex to interpret. On the other hand, if the input data is already encoded for structural information as in (Akbar et al. 2021; Robert et al. 2021), then the model only needs to learn semantic mapping rules, which might be more interpretable for rational antibody design.

Tokenization, which determines how the input sequence is subdivided, further informs the rules learned by the LM. The lexicon as described by the linguistic formalization establishes the desired characteristics for tokens, and in the case of multiple possible lexicons, the lexicon used by the target type of rules should be chosen. Thus, even if the input is encoded for structural information, the lack of tokenization based on the semantic lexicon still means that the LM has to latently learn meaningful units through complex dependencies. Thus, if the goal is to extract only compositional semantic rules from the model, the input should be encoded with structural information and be tokenized so that the target tokens map to abstract functional meaning, as characterized by the linguistic formalization.

In conclusion, a linguistic formalization, which rigorously defines biological sequences in terms of a natural language system with a lexicon and grammar, provides more explicit guidance on how specific LM design choices can affect the types of rules latently learned by the LM. For a targeted learning of only compositional semantic rules, antibody LMs should receive input that is encoded with structural information and tokenized based on a semantic lexicon. Although there is a lack of sufficiently large-scale experimental antibody-antigen binding data with structurally resolved information, simulated data can still offer an opportunity to assess the methodological proposals suggested here (Sandve and Greiff 2022; Robert et al. 2021). More generally, however, semantic tokens with known functional meaning remain unexplored, which hinders the building of more interpretable antibody LMs. Although we have here only shown a formalization of the antibody language, similar formalizations could prove invaluable for other interpretable biological sequence modeling and point to new insights into existing biological questions.

### Funding

We acknowledge generous support by The Leona M. and Harry B. Helmsley Charitable Trust (#2019PG-T1D011, to VG), UiO World-Leading Research Community (to VG), UiO:LifeScience Convergence Environment Immunolingo (to VG, GKS, and DTTH), EU Horizon 2020 iReceptorplus (#825821) (to VG), a Research Council of Norway FRIPRO project (#300740, to VG), a Research

Council of Norway IKTPLUSS project (#311341, to VG and GKS), a Norwegian Cancer Society Grant (#215817, to VG), and Stiftelsen Kristian Gerhard Jebsen (K.G. Jebsen Coeliac Disease Research Centre) (to GKS).

**Declaration of interests**

V.G. declares advisory board positions in aiNET GmbH, Enpicom B.V, Specifica Inc, Adaptyv Biosystems, EVQLV, and Omniscope. V.G. is a consultant for Roche/Genentech, immunai, and Proteinea.